\documentclass[conference]{IEEEtran}
\IEEEoverridecommandlockouts
\usepackage{cite}
\usepackage{amsmath,amssymb,amsfonts}
\usepackage{algorithmic}
\usepackage{graphicx}
\usepackage{textcomp}
\usepackage{xcolor}
\def\BibTeX{{\rm B\kern-.05em{\sc i\kern-.025em b}\kern-.08em
    T\kern-.1667em\lower.7ex\hbox{E}\kern-.125emX}}

\usepackage{braket}

\begin{document}

\title{Parallel Quantum Computing Emulation
\thanks{This work was supported by the Office of Naval Research under Grant No.\ N00014-17-1-2107.}
}

% double blind review
\author{
\IEEEauthorblockN{Brian R. La Cour}
\IEEEauthorblockA{\textit{Applied Research Laboratories} \\
\textit{The University of Texas at Austin}\\
Austin, United States \\
blacour@arlut.utexas.edu}
\and
\IEEEauthorblockN{S. Andrew Lanham}
\IEEEauthorblockA{\textit{Applied Research Laboratories} \\
\textit{The University of Texas at Austin}\\
Austin, United States \\
andrew.lanham@arlut.utexas.edu}
\and
\IEEEauthorblockN{Corey I. Ostrove}
\IEEEauthorblockA{\textit{Applied Research Laboratories} \\
\textit{The University of Texas at Austin}\\
Austin, United States \\
costrove@arlut.utexas.edu}
}

\maketitle

\begin{abstract}
Quantum computers provide a fundamentally new computing paradigm that promises to revolutionize our ability to solve broad classes of problems.  Surprisingly, the basic mathematical structures of gate-based quantum computing, such as unitary operations on a finite-dimensional Hilbert space, are not unique to quantum systems but may be found in certain classical systems as well.

Previously, it has been shown that one can represent an arbitrary multi-qubit quantum state in terms of classical analog signals using nested quadrature amplitude modulated signals.  Furthermore, using digitally controlled analog electronics one may manipulate these signals to perform quantum gate operations and thereby execute quantum algorithms.  The computational capacity of a single signal is, however, limited by the required bandwidth, which scales exponentially with the number of qubits when represented using frequency-based encoding.

To overcome this limitation, we introduce a method to extend this approach to multiple parallel signals.  Doing so allows a larger quantum state to be emulated with the same gate time required for processing frequency-encoded signals.  In the proposed representation, each doubling of the number of signals corresponds to an additional qubit in the spatial domain.  Single quit gate operations are similarly extended so as to operate on qubits represented using either frequency-based or spatial encoding schemes.  Furthermore, we describe a method to perform gate operations between pairs of qubits represented using frequency or spatial encoding or between frequency-based and spatially encoded qubits.  Finally, we describe how this approach may be extended to represent qubits in the time domain as well.
\end{abstract}
% limit of 250 words

%\begin{IEEEkeywords}
%TBD
%\end{IEEEkeywords}

%%%%%%%%%%%%%%%%%%%%%%%%%%%%%%%%%%%%%%%

\section{Introduction}

Quantum computers provide a fundamentally new computing paradigm that promises to revolutionize our ability to solve certain broad classes of problems that would be impractical to solve on a classical, digital computer \cite{Feynman1982,Deutsch1985,Mermin}.  Examples of such problems include prime factorization and unstructured searches \cite{Shor1994,Grover1997}.  The construction and operation of such devices is, however, extremely challenging, as unwanted interactions with the environment can quickly lead to decoherence and a subsequent loss of computational efficacy \cite{Mike&Ike,Joos}.  Current devices are modest in scale (i.e., less than 100 qubits) and require careful isolation from the environment, thereby limiting their practical utility \cite{Debnath2016,Barends2014,Boixo2018}.  Although there are no known theoretical limits to the scalability of such quantum devices, their practical feasibility is as yet unknown.

The basic mathematical structures of gate-based quantum computing, such as unitary operations on a finite-dimensional Hilbert space, are not unique to quantum systems but may be found in certain classical systems as well \cite{Spreeuw1998,Spreeuw2001,Dragoman,Kish2003}.  One is thus led to consider other analogue physical systems that possess similar characteristics and, so, may be useful for performing computations.  This raises the interesting question of whether such classical emulations can, indeed, be realized and, if so, what practical advantages they might offer.

Prior work has demonstrated that analog electronic signal processing can indeed provide a mathematically equivalent representation of a quantum computer and, therefore, provide an alternative physical implementation \cite{LaCour&Ott2015,LaCour2016}.  This approach to quantum emulation uses the frequency domain of a signal to encode information, and approach we shall call \emph{frequency-based encoding}.  In this scheme, an $n$-qubit quantum state is represented by the complex amplitudes of $2^n$ uniformly spaced frequencies of a given complex signal.  These, in turn, are generated by products of $n$ in-phase or quadrature signals with octavely spaced frequencies.

Due to the octave spacing of the qubit frequencies, which is necessary to avoid frequency collision or bunching in the combined signal, the required bandwidth scales exponentially with the number of qubits.  The time to perform a single gate operation is therefore limited by the lowest frequency qubit but is unchanged if higher frequency qubits are added.  For example, a signal operating in the 1 MHz to 1 GHz range can be used to represent 10 qubits, since $2^{10} \approx 1$ GHz/MHz.  Each gate operation would therefore take on the order of a microsecond to perform and, yet, would be capable of operating on all 1024 different 10-bit binary states at once, thus giving an \emph{effective} gate time on the order of a nanosecond due to the inherent parallelism of the device.  Even at this modest scale one can achieve a computational speedup up to two orders of magnitude higher than current digital processors \cite{LaCour&Ott&Lanham2017}.  To go beyond this, though, we must either increase the upper end of the frequency band or find new ways to encode and manipulate information.  The latter alternative is the primary subject of this paper.

There are several ways in which one might increase the number of qubits beyond simply increasing the bandwidth of the signal.  One straightforward approach is to use a train of signals, rather than a single signal, to represent a larger quantum state.  In this approach, which we call \emph{time-based encoding}, each doubling of the number of signals results in an additional qubit, much as the doubling of frequencies in the octave spacing scheme results in an additional qubit.  Thus, a train of $L = 2^\ell$ signals, each with $N = 2^n$ frequencies, can be used to represent a quantum state of $\ell+n$ qubits.  This approach has been introduced for use in quantum communications using photons and is easily generalized to the classical regime \cite{Humphreys2013,Donohue2013}.

Operations may be performed on individual qubits by a process of dividing, manipulating, and recombining pairs of signal trains, much as we have done in the frequency domain.  Conceptually, this process is analogous to railroad switching to rearrange a train of cars in a rail yard.  This approach has the advantage of being relatively straightforward to implement, yet it confers no particular computational speedup, as the advantage gained by the additional qubits is offset by the increased time required for each gate operation.  On the other hand, time encoded qubits may be used to implement fault-tolerant quantum error correction schemes and, thus, may find utility in other applications, such as classical communication in harsh or noisy environments.

To achieve a true speed advantage, one could instead use a set of parallel signals, each in a separate set of wires, say.  This approach, which we call \emph{spatial encoding}, is mathematically similar to time-based encoding but incurs no sacrifice in gate speed.  To achieve a true speed advantage, however, it is important that the qubits across separate signals be entangled \cite{Jozsa&Linden2003}.  Having $M = 2^m$ independent, separable signals would only confer a speed advantage of a factor of $m$.  By contrast, the ability to fully entangle both frequency and spatial qubits would confer a speed advantage of a factor of $M$.  For example, $M = 1024$ signals, each operating in a frequency range of 1 MHz to 1 GHz, can be used to encode a 20-qubit signal with an effective gate time on the order of picoseconds and, hence, an equivalent digital processor clock speed that is three orders of magnitude faster.  In accordance with Ref.\ \cite{LaCour&Ott&Lanham2017}, this would imply a speedup by up to five orders of magnitude over a modern digital processor.

The paper is organized as follows.  In Sec.\ \ref{sec:frequency} we review the frequency-based approach to quantum state representation and quantum computing emulation.  Next, in Sec.\ \ref{sec:spatial}, we describe how this approach may be extended to parallel signals.  In particular, we introduce a method for performing two-qubit gate operations between frequency-based and spatial qubits, thereby allowing for arbitrary, fully entangled states.  In Sec.\ \ref{sec:time} we provide a straightforward generalization of this approach to time-based qubits, thus permitting frequency-based, time-based, and spatial qubit to all be mutually entangled.  Finally, in Sec.\ \ref{sec:practical}, we assess the practical advantages, challenges, and limitations of the proposed physical implementation.  Our conclusions are summarized in Sec.\ \ref{sec:conclusion}.

%%%%%%%%%%%%%%%%%%%%%%%%%%%%%%%%%%%%%%%

\section{Frequency-Based Encoding}
\label{sec:frequency}

The basic mathematical structure of gate-based quantum computing is that of a tensor product of two-dimensional Hilbert spaces, commonly referred to as quantum bits or ``qubits.''  In the context of classical signal processing, there are many ways such mathematical structures may be represented.  One appealing approach is to use signals within a quadrature amplitude modulation scheme to represent a single qubit.

Consider a pair of complex, base-banded in-phase and quadrature signals $e^{j\omega_0t}$ and $e^{-j\omega_0t}$ defined for some radian frequency $\omega_0 > 0$ and over some time interval $t \in [0,T)$, where $T$ is a multiple of the fundamental period $2\pi/\omega_0$.  These two signals may be used to represent that classical bits 0 and 1.  Furthermore, we may construct complex linear combinations of these two signals.  Thus, $\psi(t) = \alpha \, e^{j\omega_0t} + \beta \, e^{-j\omega_0t}$, for $\alpha, \beta \in \mathbb{C}$, would represent a general qubit state.  (Note: We assume $|\alpha|^2 + |\beta|^2 > 0$; more commonly this sum is set to one.)  \textcolor{black}{A superposition of quantum states is represented by a superposition of signals.  Thus, the quantum superposition $\ket{\psi} = \ket{0} - \ket{1}$ is represented by the classical signal $\psi(t) = 2\cos(\omega_0t)$.}

To add additional qubits to the representation, we use in-phase and quadrature signals of different frequencies.  For $n$ such frequencies $\omega_0 < \omega_1 < \cdots < \omega_{n-1}$, we multiply the $n$ corresponding complex sinusoidal signals, either of the in-phase or quadrature variety, to obtain a signal with one of $N = 2^n$ possible frequencies, half of which will be negative.  In particular, taking an octave spacing such that $\omega_i = 2^i\omega_0$ results in sum and difference frequencies that are uniformly spaced.

Each such product corresponds to a basis function $\phi_x$, where, for $x = \sum_{i=0}^{n-1} x_i 2^i$ and $x_i \in \{0,1\}$,
\begin{equation}
\phi_x(t) = \exp\left[ \sum_{i=0}^{n-1} (-1)^{x_i} j\omega_i t\right] 1_{[0,T)}(t)
\end{equation}
and $1_{[0,T)}(t) = 1$ for $t \in [0,T)$ and is zero otherwise.

\begin{figure}
\centerline{\scalebox{0.25}{\includegraphics{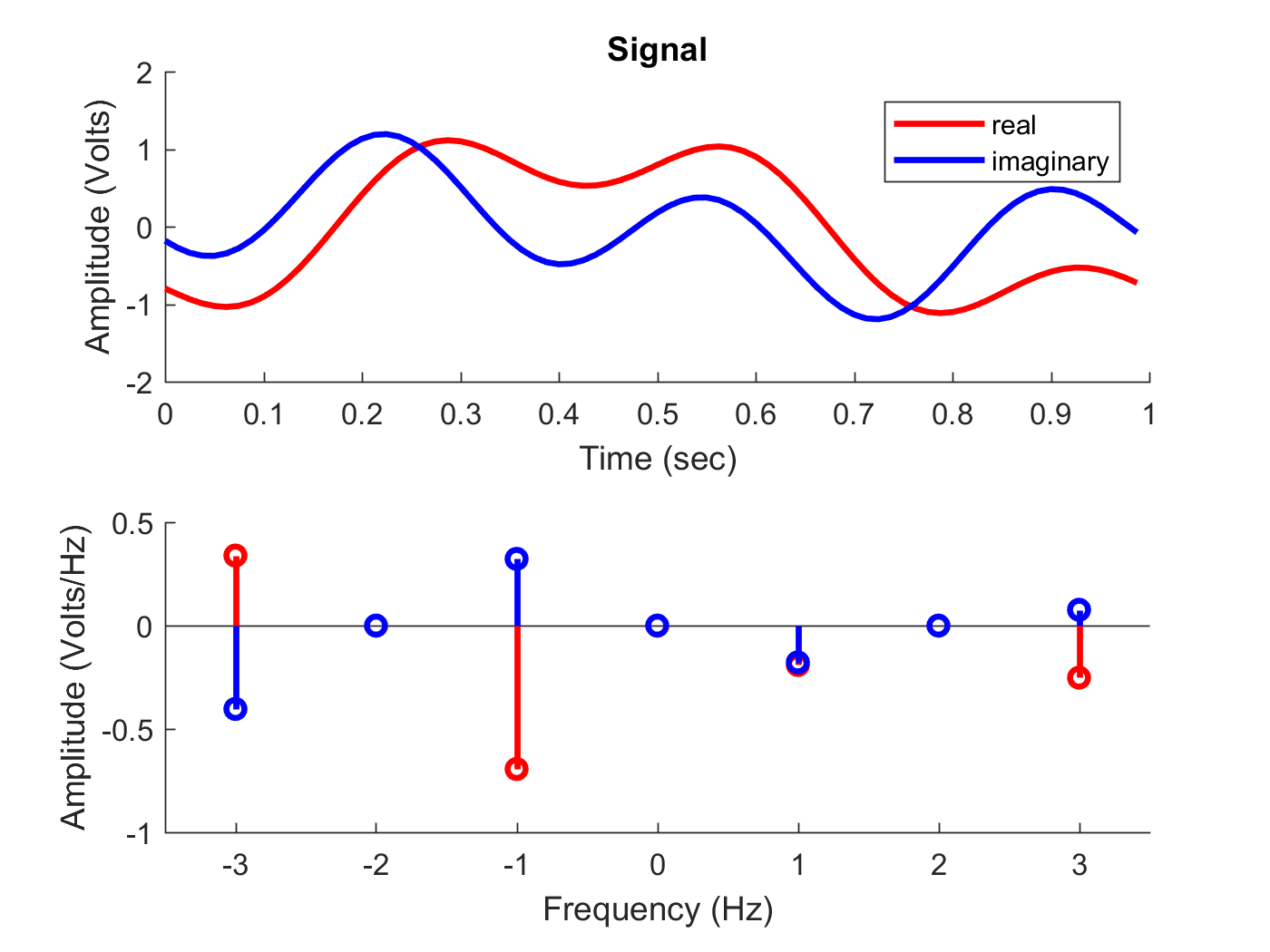}}}
\caption{\color{black}(Color online) Plot of the frequency-encoded representation of an arbitrary two-qubit state with $\alpha_0 = -0.2518 + 0.0766j$, $\alpha_1 = -0.1907 - 0.1778j$, $\alpha_2 = -0.6936 + 0.3228j$, $\alpha_3 = 0.3389 - 0.4032j$, and $\omega_0 = 1~\mathrm{Hz}$.  The top plot shows the time-domain signal, while the bottom plot shows the Fourier transform of the signal.  Note that the nonzero complex Fourier components match the four corresponding complex amplitudes $\alpha_0, \alpha_1, \alpha_2, \alpha_3$ at frequencies 3~Hz, 1~Hz, -1~Hz, -3~Hz, respectively.\color{black}}
\label{fig:two-qubit}
\end{figure}

A general $n$-qubit state may then be represented by a complex linear combination of these $N$ basis functions such that
\begin{equation}
\psi(t) = \sum_{x=0}^{N-1} \alpha_x \phi_x(t) \; ,
\end{equation}
where $\alpha_x \in \mathbb{C}$.  \color{black}Note that this general form is capable of representing any $n$-qubit quantum state, including entangled (i.e., nonseparable) quantum states.  For example, the Bell state $\ket{\psi} = \ket{00} - \ket{11}$, a maximally entangled two-qubit state, would be represented by the signal $\psi(t) = 2\sin(3\omega_0t)$.  A general two-qubit state would be represented by a signal of the form
\begin{equation}
\psi(t) = \alpha_0 e^{j3\omega_0t} + \alpha_1 e^{j\omega_0t} + \alpha_2 e^{-j\omega_0t} + \alpha_3 e^{-j3\omega_0t} \; .
\label{eqn:two-qubit}
\end{equation}
A particular example is illustrated in Fig.\ \ref{fig:two-qubit}.\color{black}

Finally, we define an inner product between two such signals, $\psi$ and $\varphi$, as follows:
\begin{equation}
\braket{\varphi|\psi} = \frac{1}{T} \int_{0}^{T} \varphi(t)^* \psi(t) \, dt \; .
\end{equation}
Note that, in particular, this implies $\braket{\phi_x|\psi} = \alpha_x$.  Physically, this corresponds to the multiplication of two complex signals, the product of which is passed through a low-pass filter.

The inner product lends to a description in terms of the bra-ket notation commonly used in quantum computing.  Thus, the ``ket'' $\ket{\psi}$ is equated to the signal $\psi$, and the ``bra'' $\bra{\varphi}$ is the linear operator (i.e., low-pass filter) that maps $\psi$ to $\braket{\varphi|\psi}$.  Similarly, the basis function $\phi_x$ is equated to the basis ket $\ket{x}$, so that $\ket{\psi}$ may be written
\begin{equation}
\ket{\psi} = \sum_{x=0}^{N-1} \ket{x}\braket{x|\psi} = \sum_{x=0}^{N-1} \alpha_x \ket{x} \; .
\end{equation}
For qubit $i$, the basis kets $\ket{0}_i, \ket{1}_i$ represent the basis signals $e^{j\omega_it}$ and $e^{-j\omega_it}$.  The tensor product of these, denoted $\ket{x_{n-1}} \otimes \cdots \otimes \ket{x_0}$, is the basis ket $\ket{x}$ and corresponds to the basis signal $\phi_x$.  Note that the tensor product here is just functional multiplication and will be omitted in what follows.

%-----------------------------------------------------------------------------------

\subsection{Projections}

Project operations are used to decompose signals into qubit-specific subspaces for gate operation and measurement.  For a projection on the two subspaces of qubit $i$, we use the notation
\begin{equation}
\ket{\psi} = \Pi_0^{(i)} \ket{\psi} + \Pi_1^{(i)} \ket{\psi} \; ,
\end{equation}
where $\Pi_0^{(i)} \ket{\psi}$ denotes a signal with frequencies $\sum_{i=0}^{n-1} (-1)^{x_i} \omega_i$ such that $x_i = 0$ and $\Pi_0^{(i)} \ket{\psi}$ denotes a signal with frequencies $\sum_{i=0}^{n-1} (-1)^{x_i} \omega_i$ such that $x_i = 1$.  \color{black}For the two-qubit state of Eqn.\ (\ref{eqn:two-qubit}), the projections for qubit $i=1$ are $\Pi_0^{(i)}\ket{\psi} = \alpha_0\ket{00} + \alpha_1\ket{01}$ and $\Pi_1^{(i)}\ket{\psi} = \alpha_2\ket{10} + \alpha_3\ket{11}$, which are represented by the signals
\begin{align}
\Pi_0^{(i)}\ket{\psi} \; &\longleftrightarrow \; e^{j2\omega_0t} (\alpha_0 e^{j\omega_0t} + \alpha_1 e^{-j\omega_0t}) \\
\Pi_1^{(i)}\ket{\psi} \; &\longleftrightarrow \; e^{-j2\omega_0t} (\alpha_2 e^{j\omega_0t} + \alpha_3 e^{-j\omega_t}) \; ,
\end{align}
respectively.  \color{black}Physically, these projected signals are produced using comb-like filters as described in \cite{LaCour&Ott2015}.

The filtering process actually produces four signals: the so-called partial projection signals $\ket{\psi_0^{(i)}}, \ket{\psi_1^{(i)}}$, and the single-qubit signals $\ket{0}_i, \ket{1}_i$.  These are related to the projection as follows:
\begin{align}
\Pi_0^{(i)} \ket{\psi} &= \ket{0}_i \ket{\psi_0^{(i)}} \\
\Pi_1^{(i)} \ket{\psi} &= \ket{1}_i \ket{\psi_1^{(i)}}
\end{align}
Thus, the projection corresponds to the decomposition
\begin{equation}
\psi(t) = e^{j\omega_it} \psi_0^{(i)}(t) + e^{-j\omega_it} \psi_0^{(i)}(t) \; .
\end{equation}
\color{black}For the two-qubit state of Eqn.\ (\ref{eqn:two-qubit}), the partial projections for qubit $i=1$ would be $\psi_0^{(i)}(t) = \alpha_0 e^{j\omega_0t} + \alpha_1 e^{-j\omega_0t}$ and $\psi_1^{(i)}(t) = \alpha_2 e^{j\omega_0t} + \alpha_3 e^{-j\omega_0t}$.  \color{black}

Projections into four subspaces corresponding to two qubits works similarly.  Thus, $\ket{\psi}$ is decomposed as follows:
\begin{equation}
\ket{\psi} = \Pi_{00}^{(ij)} \ket{\psi} + \Pi_{01}^{(ij)} \ket{\psi} + \Pi_{10}^{(ij)} \ket{\psi} + \Pi_{11}^{(ij)} \ket{\psi} \; .
\end{equation}
Each such projection may be formed from two single-qubit projections in a manner described above and will result in a set of three signals such that
\begin{align}
\Pi_{00}^{(ij)} \ket{\psi} &= \ket{0}_i \ket{0}_j \ket{\psi_{00}^{(ij)}} \\
\Pi_{01}^{(ij)} \ket{\psi} &= \ket{0}_i \ket{1}_j \ket{\psi_{01}^{(ij)}} \\
\Pi_{10}^{(ij)} \ket{\psi} &= \ket{1}_i \ket{0}_j \ket{\psi_{00}^{(ij)}} \\
\Pi_{11}^{(ij)} \ket{\psi} &= \ket{1}_i \ket{1}_j \ket{\psi_{01}^{(ij)}}
\end{align}
and corresponding to the signal decomposition
\begin{equation}
\begin{split}
\psi(t) &= e^{j(\omega_i+\omega_j)t} \psi_{00}^{(ij)}(t) + e^{j(\omega_i-\omega_j)t} \psi_{01}^{(ij)}(t) \\
&+ e^{j(-\omega_i+\omega_j)t} \psi_{10}^{(ij)}(t) + e^{j(-\omega_i-\omega_j)t} \psi_{11}^{(ij)}(t) \; .
\end{split}
\end{equation}

%-----------------------------------------------------------------------------------

\subsection{Single-Qubit Gate Operations}

Once a projective decomposition has been performed, single-gate operations are quite straightforward.  A gate $U$ is a linear operator, typically unitary, that may be represented by a complex $2\times2$ matrix
\begin{equation}
U = 
\begin{pmatrix}
U_{00} & U_{01} \\ U_{10} & U_{11}
\end{pmatrix}
\end{equation}
such that
\begin{align}
U\ket{0} &= U_{00} \ket{0} + U_{10} \ket{1} \\
U\ket{1} &= U_{01} \ket{0} + U_{11} \ket{1} \; .
\end{align}
Let $U_i$ denote the single-qubit gate operation $U$ acting on qubit $i$.  The transformed quantum state/signal is then
\begin{equation}
U_i \ket{\psi} = \Bigl( U \ket{0}_i \Bigr) \otimes \ket{\psi_0^{(i)}} + \Bigl( U \ket{1}_i \Bigr) \otimes \ket{\psi_1^{(i)}}
\end{equation}
The corresponding transformed signal $\psi'$ is therefore
\begin{equation}
\begin{split}
\psi'(t) &= \Bigl( U_{00} \, e^{j\omega_it} + U_{10} \, e^{-j\omega_it} \Bigr) \psi_0^{(i)}(t) \\
&+ \Bigl( U_{01} \, e^{j\omega_it} + U_{11} \, e^{-j\omega_it} \Bigr) \psi_1^{(i)}(t) \; .
\end{split}
\end{equation}
In this manner, any single-qubit gate operation may be implemented on a classical signal.

\begin{figure}
\centerline{\scalebox{0.25}{\includegraphics{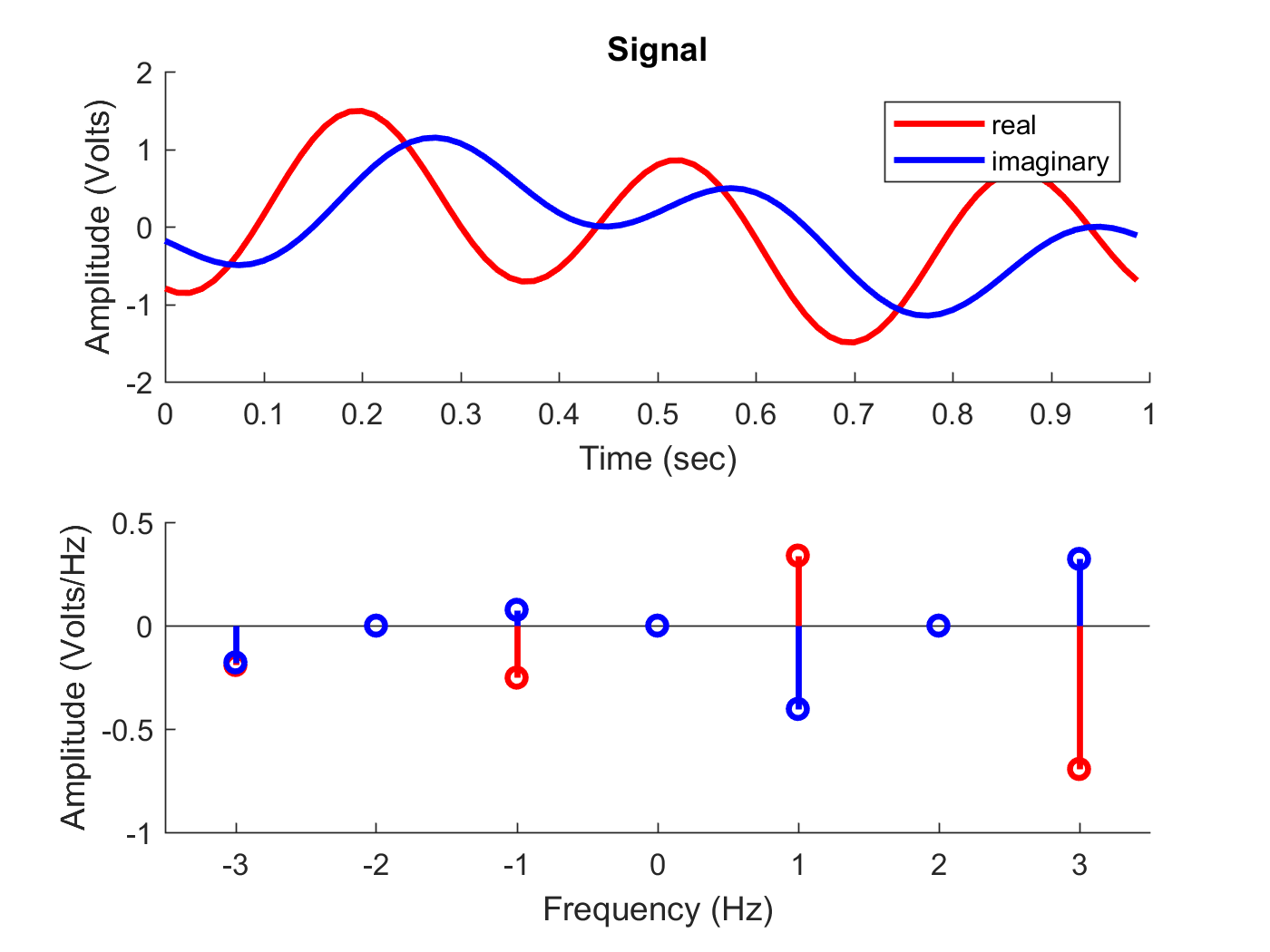}}}
\caption{\color{black}(Color online) Plot of emulated quantum state from Fig.\ \ref{fig:two-qubit} with a NOT gate applied to qubit 1.  Note that the complex amplitudes for 1, 3~Hz and -1, -3~Hz have been interchanged.\color{black}}
\label{fig:NOTgate}
\end{figure}

\color{black}For example, a NOT gate would be represented by the unitary matrix
\begin{equation}
U = \begin{pmatrix} 0 & 1 \\ 1 & 0 \end{pmatrix} \; .
\end{equation}
Applying this gate to qubit $i = 1$ for the two-qubit state of Eqn.\ (\ref{eqn:two-qubit}), we obtain
\begin{equation}
\begin{split}
\psi'(t) &= e^{-j2\omega_0t} \left( \alpha_0 e^{j\omega_0t} + \alpha_1 e^{-j\omega_0t} \right) \\
&\quad + e^{j2\omega_0t} \left( \alpha_2 e^{j\omega_0t} + \alpha_3 e^{-j\omega_0t} \right) \\
&= \alpha_0 e^{-j\omega_0t} + \alpha_1 e^{-j3\omega_0t} + \alpha_2 e^{j3\omega_0t} + \alpha_3 e^{j\omega_0t} \; .
\end{split}
\end{equation}
Note that this corresponds to the quantum state $\ket{\psi'} = \alpha_0 \ket{10} + \alpha_1 \ket{11} + \alpha_2 \ket{00} + \alpha_3 \ket{01}$, which matches $\ket{\psi}$ with a NOT gate applied to qubit 1 (i.e., the left qubit).  This is illustrated in Fig.\ \ref{fig:NOTgate} for the example from Fig.\ \ref{fig:two-qubit}.
\color{black}

%-----------------------------------------------------------------------------------

\subsection{Two-Qubit Gate Operations}

Two-qubit gate operations work in much the same manner as single-qubit opertations.  Of particular interest are controlled-$U$ operations, denoted $C_{ij}(U)$, where the single-qubit gate $U$ is applied to qubit $j$ if qubit $i$ is 1.  Thus,
\begin{align}
C_{ij}(U) \ket{0}_i \ket{0}_j &= \ket{0}_i \ket{0}_j \\
C_{ij}(U) \ket{0}_i \ket{1}_j &= \ket{0}_i \ket{1}_j \\
C_{ij}(U) \ket{1}_i \ket{0}_j &= \ket{1}_i \Bigl( U_{00} \ket{0}_j + U_{10} \ket{1}_j \Bigr) \\
C_{ij}(U) \ket{1}_i \ket{1}_j &= \ket{1}_i \Bigl( U_{01} \ket{0}_j + U_{11} \ket{1}_j \Bigr) \; .
\end{align}

Thus, operating $C_{ij}(U)$ on the $n$-qubit state $\ket{\psi}$ results in the transformed signal
\begin{equation}
\begin{split}
\psi'(t) &= e^{j\omega_it} \psi_0^{(i)}(t) \\
&\quad + e^{-j\omega_it} \Bigl( U_{00} \, e^{j\omega_jt} + U_{10} \, e^{-j\omega_jt} \Bigr) \psi_{10}^{(ij)}(t) \\
&\quad + e^{-j\omega_it}\Bigl( U_{01} \, e^{j\omega_it} + U_{11} \, e^{-j\omega_jt} \Bigr) \psi_{11}^{(ij)}(t) \; .
\end{split}
\end{equation}
The ability to perform arbitrary single-qubit and controlled two-qubit gate operations allows for universal quantum computing \cite{DiVincenzo1995}.

%-----------------------------------------------------------------------------------

\color{black}
\subsection{Measurement Gates}

Once the final, transformed state has been produced, a sequence of measurement gates may be applied to extract a digital answer.  Measurement gates are realized through single-qubit subspace projections, much as for unitary gates.  Thus, given a final state $\psi'$, we perform a measurement on qubit $i$ by filtering the signal to obtain ${\psi'_0}^{(i)}$ and ${\psi'_1}^{(i)}$.  The RMS values, $\bar{v}_0$ and $\bar{v}_1$, of these two signals are used to probabilistically select a result, with outcome $x \in  \{0, 1\}$ occurring with probability $\bar{v}_x^2/(\bar{v}_0^2 + \bar{v}_1^2)$, thereby reproducing the quantum mechanical Born rule for statistical outcomes.  (Other schemes, such as selecting the larger of $\bar{v}_0$ and $\bar{v}_1$, can also be used.)  Given an outcome $x$, the state collapses to the $(n-1)$-qubit projected state ${\psi'_x}^{(i)}$.  The process is repeated to measure all $n$ qubits, resulting in an $n$-bit digital outcome.
\color{black}

%%%%%%%%%%%%%%%%%%%%%%%%%%%%%%%%%%%%%%%

\section{Spatial Encoding}
\label{sec:spatial}

The frequency-based encoding scheme may be parallelized by considering a cluster of $M$ complex signals (represented by $2M$ real signals), each composed of $N$ frequencies representing $n$ qubits.  We suppose $M = 2^m$, where $m \ge 0$ is an integer.  Mathematically, this cluster of signals may be represented by a vector $\vec{\Psi}$ of $M$ signals, where
\begin{equation}
\vec{\Psi}(t) = \sum_{y=0}^{M-1} \psi_y(t) \, \vec{e}_y \; ,
\end{equation}
\begin{equation}
\psi_y(t) = \sum_{x=0}^{N-1} \alpha_{x,y} \, \phi_x(t) \; ,
\end{equation}
and $\vec{e}_y$ is a column vector that is 1 in row $y$ and zero elsewhere.

The Hilbert space description is completed by defining an inner product between two such vectors, $\vec{\Psi}'$ and $\vec{\Psi}$, as follows:
\begin{equation}
\braket{\vec{\Psi}'|\vec{\Psi}} = \sum_{y=0}^{M-1} \braket{\psi'_y|\psi_y} \; .
\end{equation}

In the language of quantum computing, we may write $\vec{\Psi} = \ket{\Psi}$ and $\vec{e}_y = \ket{y}$.  A tensor product of basis states may then be defined such that $\phi_x \, \vec{e}_y = \ket{x} \otimes \ket{y}$.  Note that, although $\phi_x \, \vec{e}_y = \vec{e}_y \, \phi_x$, we shall adopt the convention for the tensor product that each basis vector $\ket{x} \otimes \ket{y}$ is composed of $n$ frequency qubits (on the left) and $m$ spatial qubits (on the right).  In what follows, we shall omit the explicit use of tensor notation and simply write $\ket{x} \otimes \ket{y} = \ket{x} \ket{y}$.

%-----------------------------------------------------------------------------------

\subsection{Projections}

As with frequency-based encoding, projection operations are needed to realize gate operations on spatially encoded qubits.  Notionally, we may decompose a state $\ket{\Psi}$ as follows:
\begin{equation}
\ket{\Psi} = \sum_{\bar{y}_i} \ket{\psi_{(0,\bar{y}_i)}} \ket{0,\bar{y}_i} + \sum_{\bar{y}_i} \ket{\psi_{(1,\bar{y}_i)}} \ket{1,\bar{y}_i} \; ,
\end{equation}
where $\bar{y}_i = (y_0, y_1, \cdots y_{i-1}, y_{i+1}, \cdots y_{m-1})$ denote the binary values of $y$ for all but the $i^{\rm th}$ bit.  With this notation, we write $y = (y_i, \bar{y}_i)$ for the entire $m$-bit index $y$.

In the spatial domain, such projections are performed through switches.  Physically, this may be done through a sequence of pairwise swaps.  There will be up to $m$ such sequences, one for each spatial qubit to addressed.  For each such sequence, up to $M/2-1$ stages of swaps will be needed.  Thus, the number of swaps scales exponentially with $m$.  Note also that, once a gate operation has been performed, the signals must be reordered back to their original sequencing.

Performing projections over pairs of qubits works in a similar manner and may be accomplished by staging pairs of single-qubit projections.  Formally, we write
\begin{equation}
\begin{split}
\ket{\Psi} &= \sum_{\bar{y}_{ij}} \ket{\psi_{(0,0,\bar{y}_{ij})}} \ket{0,0,\bar{y}_{ij}} + \sum_{\bar{y}_{ij}} \ket{\psi_{(0,1,\bar{y}_{ij})}} \ket{0,1,\bar{y}_{ij}} \\
&+ \sum_{\bar{y}_{ij}} \ket{\psi_{(1,0,\bar{y}_{ij})}} \ket{1,0,\bar{y}_{ij}} + \sum_{\bar{y}_{ij}} \ket{\psi_{(1,1,\bar{y}_{ij})}} \ket{1,1,\bar{y}_{ij}} \; ,
\end{split}
\end{equation}
where, much as before, $\bar{y}_{ij}$ is the sequence of binary values of $y$ with $y_i$ and $y_j$ removed, and $y = (y_i, y_j, \bar{y}_{ij})$.  Note that, given $y_i$ and $y_j$, each $\ket{\psi_{(y_i,y_j,\bar{y}_{ij})}}$ is one of $M/4$ signals for which the $i^{\rm th}$ bit of $y$ is $y_i$ and the $j^{\rm th}$ of $y$ is $y_j$.

%-----------------------------------------------------------------------------------

\subsection{Single-Qubit Gate Operations}

Using the projection method described above, we may apply a single-qubit gate operator $U$ on spatial qubit $i$ as follows.  First, note that
\begin{equation}
U_{,i} \ket{\Psi} = \sum_{\bar{y}_i} \Bigl[ \ket{\psi_{(0,\bar{y}_i)}} U_i \ket{0,\bar{y}_i} + \ket{\psi_{(1,\bar{y}_i)}} U_i \ket{1,\bar{y}_i} \Bigr] \; ,
\end{equation}
where $U_{,i}$ is understood to mean that $U$ is operating on spatial qubit $i$.  Now, note that, formally,
\begin{align}
U_i \ket{0,\bar{y}_i} &= U_{00} \ket{0,\bar{y}_i} + U_{10} \ket{1,\bar{y}_i} \\
U_i \ket{1,\bar{y}_i} &= U_{01} \ket{0,\bar{y}_i} + U_{11} \ket{1,\bar{y}_i} \; .
\end{align}
Substituting and rearranging terms, we find
\begin{equation}
\begin{split}
U_{,i} \ket{\Psi} 
&= \sum_{\bar{y}_i} \Bigl( U_{00} \ket{\psi_{(0,\bar{y}_i)}} + U_{01} \ket{\psi_{(1,\bar{y}_i)}} \Bigr) \ket{0,\bar{y}_i} \\
&+ \sum_{\bar{y}_i} \Bigl( U_{10} \ket{\psi_{(0,\bar{y}_i)}} + U_{11} \ket{\psi_{(1,\bar{y}_i)}} \Bigr) \ket{1,\bar{y}_i} \; .
\end{split}
\end{equation}
Thus, if $\ket{\Psi'} = U_{,i} \ket{\Psi} $, then
\begin{align}
\psi'_{(0,\bar{y}_i)}(t) &= U_{00} \, \psi_{(0,\bar{y}_i)}(t) + U_{01} \, \psi_{(1,\bar{y}_i)}(t) \\
\psi'_{(1,\bar{y}_i)}(t) &= U_{10} \, \psi_{(0,\bar{y}_i)}(t) + U_{11} \, \psi_{(1,\bar{y}_i)}(t) \; .
\end{align}

\begin{figure}
\centerline{\scalebox{0.4}{\includegraphics{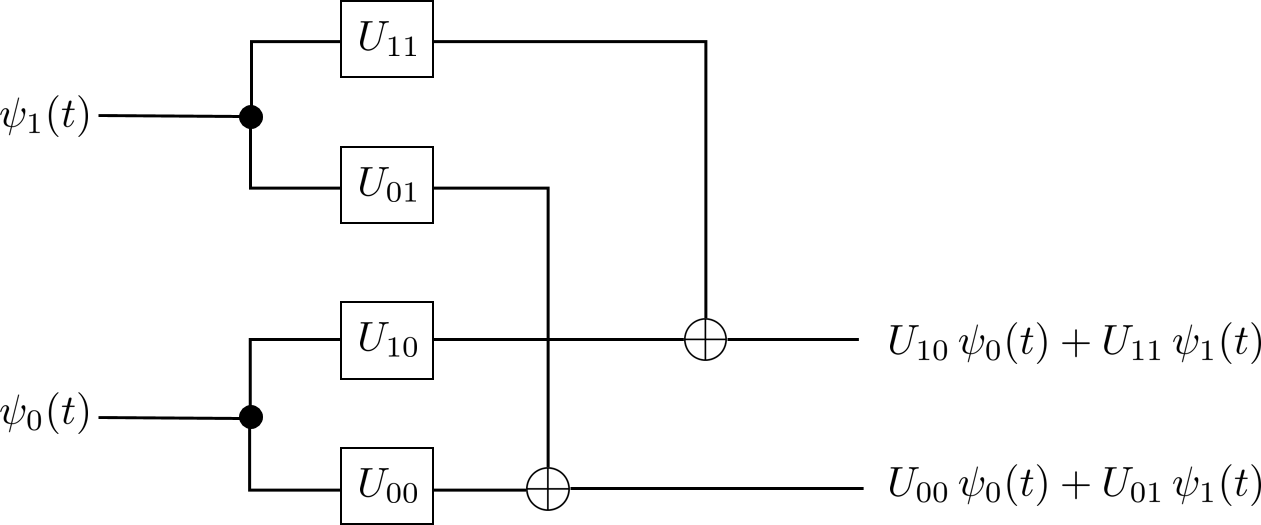}}}
\caption{Notional wire schematic for operation on a single spatial qubit for $M = 2$.  The boxes represent complex scalar multiplication, while the $\oplus$ symbol represents complex addition.}
\label{fig:tree2}
\end{figure}

The sum is only formal: the $M$ signals are, in fact, distinct.  To apply the single-qubit gate $U$ to right qubit $i$, we must separate the signals into two sets: those for which $y_i = 0$ and those for which $y_i = 1$.  For each of the $M/2$ values of $\bar{y}_i$, we make two copies of $\ket{\psi_{(0,\bar{y}_i)}}$ and $\ket{\psi_{(1,\bar{y}_i)}}$ each.  The appropriate matrix elements are multiplied to each of the four signals, and then the pairs are summed to form the replacement signals for $y = (0,\bar{y}_i)$ and $y = (1,\bar{y}_i)$.  Figure \ref{fig:tree2} illustrates the operation of the single-qubit gate operation for a single spatial qubit (i.e., $M = 2$).

By contrast, applying $U$ to frequency qubit $i$ is denoted by
\begin{equation}
U_{i,} \ket{\Psi} = \sum_{y} U_{i,} \ket{\psi_y} \ket{y} = \sum_{y} \Bigl( U_i \ket{\psi_y} \Bigr) \ket{y} \; ,
\end{equation}
where $U_i \ket{\psi_y}$ is defined as before.  Note that $U_{i,}$ applies the same operation to all $M$ elements of $\ket{\Psi}$; whereas, $U_{,i}$ applies $U$ across pairs of signals.

%-----------------------------------------------------------------------------------

\subsection{Two-Qubit Gate Operations}

Performing a two-qubit operation on a pair of spatial qubits works in much the same way as that for single qubits.  First, we reorder the signals to group the four values of $(y_i, y_j, \bar{y}_{ij})$ for each $\bar{y}_{ij}$.  Suppose we perform a controlled-$U$ operation $C_{,ij}(U)$ with spatial qubit $i$ as the control and spatial qubit $j$ as the target.  Noting that
\begin{equation}
\begin{split}
\ket{\Psi} &= \sum_{\bar{y}_j} \ket{\psi_{(0,\bar{y}_j)}} \ket{0,\bar{y}_j} \\
&+ \sum_{\bar{y}_{ij}} \Bigl[ \ket{\psi_{(1,0,\bar{y}_{ij})}} \ket{1,0,\bar{y}_{ij}} + \ket{\psi_{(1,1,\bar{y}_{ij})}} \ket{1,1,\bar{y}_{ij}} \Bigr] \; ,
\end{split}
\end{equation}
we have
\begin{equation}
\begin{split}
C_{,ij}&(U) \ket{\Psi} \\
&= \sum_{\bar{y}_j} \ket{\psi_{(0,\bar{y}_j)}} \ket{0,\bar{y}_j} \\
&+ \sum_{\bar{y}_{ij}} \ket{\psi_{(1,0,\bar{y}_{ij})}} \Bigl( U_{00} \ket{1,0,\bar{y}_{ij}} + U_{10} \ket{1,1,\bar{y}_{ij}} \Bigr) \\
&+ \sum_{\bar{y}_{ij}} \ket{\psi_{(1,1,\bar{y}_{ij})}} \Bigl( U_{01} \ket{1,0,\bar{y}_{ij}} + U_{11} \ket{1,1,\bar{y}_{ij}} \Bigr) \; .
\end{split}
\end{equation}
Upon rearranging terms, we find
\begin{equation}
\begin{split}
C_{,ij}&(U) \ket{\Psi} \\
&= \sum_{\bar{y}_j} \ket{\psi_{(0,\bar{y}_j)}} \otimes \ket{0,\bar{y}_j} \\
&+ \sum_{\bar{y}_{ij}} \Bigl( U_{00} \ket{\psi_{(1,0,\bar{y}_{ij})}} + U_{01} \ket{\psi_{(1,1,\bar{y}_{ij})}} \Bigr) \ket{1,0,\bar{y}_{ij}} \\
&+ \sum_{\bar{y}_{ij}} \Bigl( U_{10} \ket{\psi_{(1,0,\bar{y}_{ij})}} + U_{11} \ket{\psi_{(1,1,\bar{y}_{ij})}} \Bigr) \ket{1,1,\bar{y}_{ij}} \; .
\end{split}
\end{equation}

Finally, we may consider a controlled-$U$ operation $C_{i,j}(U)$ with frequency qubit $i$ and spatial qubit $j$.  In this case, we first note that $\ket{\Psi}$ may be decomposed as follows:
\begin{equation}
\begin{split}
\ket{\Psi} &= \sum_{y} \Pi_0^{(i)} \ket{\psi_y} \ket{y} \\
&+ \sum_{\bar{y}_j} \Bigl[ \Pi_1^{(i)} \ket{\psi_{(0,\bar{y}_j)}} \ket{0,\bar{y}_j} + \sum_{\bar{y}_j} \Pi_1^{(i)} \ket{\psi_{(1,\bar{y}_j)}} \ket{1,\bar{y}_j} \Bigr] \; .
\end{split}
\end{equation}
Application of $C_{i,j}(U)$ therefore yields
\begin{equation}
\begin{split}
C_{i,j}&(U) \ket{\Psi} \\
&= \sum_{y} \Pi_0^{(i)} \ket{\psi_y} \ket{y} \\
&+ \sum_{\bar{y}_j} \Pi_1^{(i)} \ket{\psi_{(0,\bar{y}_j)}} \Bigl( U_{00} \ket{0,\bar{y}_j} + U_{10} \ket{1,\bar{y}_j} \Bigr) \\
&+ \sum_{\bar{y}_j} \Pi_1^{(i)} \ket{\psi_{(1,\bar{y}_j)}} \Bigl( U_{01} \ket{0,\bar{y}_j} + U_{11} \ket{1,\bar{y}_j} \Bigr) \\
&= \sum_{\bar{y}_j} \Pi_0^{(i)} \ket{\psi_{(0,\bar{y}_j)}} \ket{0,\bar{y}_j} + \sum_{\bar{y}_j} \Pi_0^{(i)} \ket{\psi_{(1,\bar{y}_j)}} \ket{1,\bar{y}_j} \\
&+ \sum_{\bar{y}_j} \left( U_{00} \Pi_1^{(i)} \ket{\psi_{(0,\bar{y}_j)}} + U_{01} \Pi_1^{(i)} \ket{\psi_{(1,\bar{y}_j)}} \right) \ket{0,\bar{y}_j} \\
&+ \sum_{\bar{y}_j} \left( U_{10} \Pi_1^{(i)} \ket{\psi_{(0,\bar{y}_j)}} + U_{11} \Pi_1^{(i)} \ket{\psi_{(1,\bar{y}_j)}} \right) \ket{1,\bar{y}_j}
\end{split}
\end{equation}

Finally, we may consider the controlled-$U$ operator $\overline{C}_{i,j}(U)$ for which spatial qubit $j$ is the control and frequency qubit $i$ is the target.  Using the spatial projection, this is simply
\begin{equation}
\overline{C}_{i,j}(U) \ket{\Psi} = \sum_{\bar{y}_j} \ket{\psi_{(0,\bar{y}_j)}} \ket{0,\bar{y}_j} 
+ \sum_{\bar{y}_j} U_i \ket{\psi_{(1,\bar{y}_j)}} \ket{1,\bar{y}_j} \; ,
\end{equation}
where $U_i$ acts on the $i^{\rm th}$ frequency qubit of $\ket{\psi_{(1,\bar{y}_j)}}$, as described previously.

%%%%%%%%%%%%%%%%%%%%%%%%%%%%%%%%%%%%%%%

\section{Time-Based Encoding}
\label{sec:time}

The extension of spatial encoding to time-based encoded is fairly straightforward.  In this scheme, $\ell$ qubits may be represented by a wavetrain of $L = 2^\ell$ signals, each of which is a collection of $M$ parallel signals with $N$ distinct frequencies.  We write this as follows:
\begin{equation}
\vec{\boldsymbol{\Psi}}(t) = \sum_{x=0}^{N-1} \sum_{y=0}^{M-1} \sum_{z=0}^{L-1} \alpha_{x,y,z} \, \phi_x(t-zT) \, \vec{e}_y \; .
\end{equation}

In bra-ket notation, we shall equate $\ket{\boldsymbol{\Psi}}$ to $\vec{\boldsymbol{\Psi}}$ and write
\begin{equation}
\ket{\boldsymbol{\Psi}} = \sum_{x=0}^{N-1} \sum_{y=0}^{M-1} \sum_{z=0}^{L-1} \alpha_{x,y,z} \ket{x} \ket{y} \otimes \ket{z} \; ,
\end{equation}
where $z = \sum_{k=0}^{\ell-1} z_k 2^k$ is the decimal form of the binary sequence $[z_{\ell-1} \cdots z_1 z_0]$.  The tensor operator $\otimes$ above is defined by a left-acting shift operator $S_z$ such that
\begin{equation}
\ket{x}\ket{y}\otimes\ket{z} := S_z \ket{x}\ket{y}
\end{equation}
and, for any $\ket{x} = \phi_x$,
\begin{equation}
(S_z\phi_x)(t) := \phi_x(t - zT) \; .
\end{equation}

Note that the time basis state $\ket{z} = \ket{z_{\ell-1} \cdots z_1 z_0}$ may be thought of as a composition of single-qubit time basis states, since
\begin{equation}
S_z = S_{2^{\ell-1}}^{z_{\ell-1}} \circ \cdots \circ S_{2^1}^{z_1} \circ S_{2^0}^{z_0} \; ,
\end{equation}
where $S^1_z = S_z$ and $S^0_z = 1$ is the identity operator.  Since $S_z$ is a linear operator, this composition of shift operators is a bilinear operation and therefore also constitutes a valid tensor product.

%-----------------------------------------------------------------------------------

\subsection{Projections}

As for spatial encoding, we may formally write a wavetrain of $M$ parallel signals, each composed of $N$ distinct frequencies, as follows:
\begin{equation}
\ket{\boldsymbol{\Psi}} = \sum_{z=0}^{L-1} \ket{\Psi_z} \otimes \ket{z}
\end{equation}
As before, this may be formally decomposed as follows:
\begin{equation}
\ket{\boldsymbol{\Psi}} = \sum_{\bar{z}_i} \ket{\Psi_{(0,\bar{z}_i)}} \otimes \ket{0,\bar{z}_i} + \sum_{\bar{z}_i} \ket{\Psi_{(1,\bar{z}_i)}} \otimes \ket{1,\bar{z}_i} \; ,
\end{equation}
where $\bar{z}_i = (z_0, z_1, \ldots, z_{i-1}, z_{i+1}, \ldots, z_{\ell-1})$.  The decomposition into four qubit subspaces is written similarly.

%-----------------------------------------------------------------------------------

\subsection{Single-Qubit Gate Operations}

Physically, projection operations are performed with a combination of switches and delays, much like to rearrangement of box cars at a train depot.  Unlike actual trains, however, the wavetrains are split, and thereby copied, to perform gate operations.  Thus, a wavetrain $\vec{\boldsymbol{\Psi}}(t) = \vec{\Psi}_0(t) + \vec{\Psi}_1(t-T)$ consisting of just one time qubit would be physically split to form two identical wavetrains $[\vec{\boldsymbol{\Psi}}(t); \vec{\boldsymbol{\Psi}}(t)]$.

\begin{figure}
\centerline{\scalebox{0.5}{\includegraphics{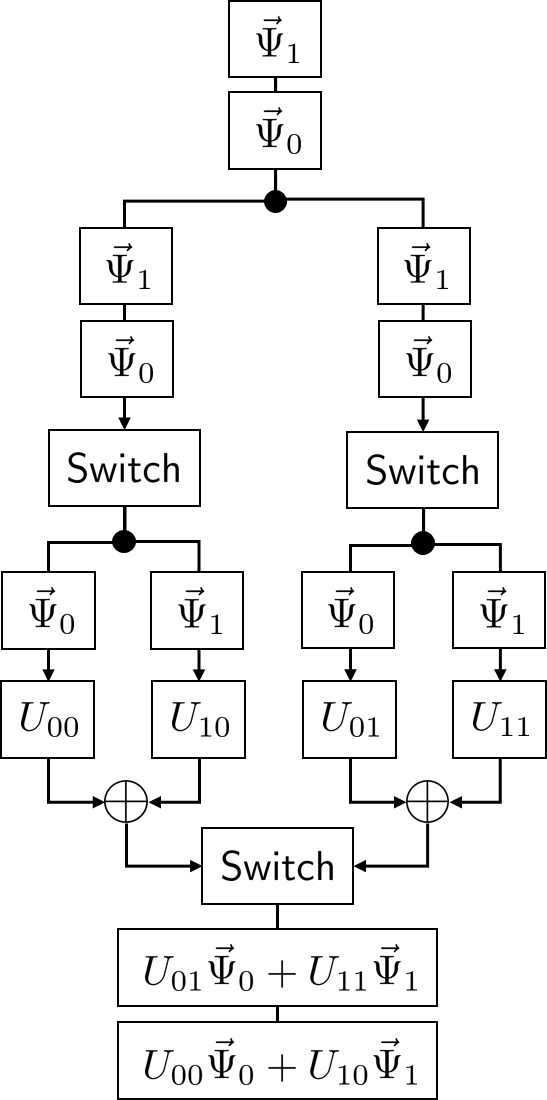}}}
\caption{Notional wire diagram for operation on a single time qubit for $L = 2$.  The boxes containing $\vec{\Psi}_0$ and $\vec{\Psi}_1$ represent time-delayed signals.  The boxes denoted \textsf{Switch} serve to separate and time delay the 0 and 1 components.}
\label{fig:train}
\end{figure}

A gate operation $U$ on this time qubit would be performed by combining the components of each of the two wavetrains, as illustrated in Fig.\ \ref{fig:train}.  First, we route the four components to four separate tracks to obtain $[\vec{\Psi}_0(t), \vec{\Psi}_1(t-T)$; $\vec{\Psi}_0(t)$, $\vec{\Psi}_1(t-T)]$ and delay two of the four so that we now have, modular an overall delay, $[\vec{\Psi}_0(t)$, $\vec{\Psi}_1(t)$; $\vec{\Psi}_0(t)$, $\vec{\Psi}_1(t)]$.  Next, we apply the gate $U$ to the first and second pair using scalar multiplication and addition.  We now have two sets of signals: $U_{00} \vec{\Psi}_0(t) + U_{10} \vec{\Psi}_1(t)$ and $U_{01} \vec{\Psi}_0(t) + U_{11} \vec{\Psi}_1(t)$.  If we now time delay the latter relative to the former and combine them, we obtain the transformed state $(U_{,,i}\vec{\boldsymbol{\Psi}})(t)$ given by
\begin{equation}
U_{00} \vec{\Psi}_0(t) + U_{10} \vec{\Psi}_1(t) + U_{01} \vec{\Psi}_0(t-T) + U_{11} \vec{\Psi}_1(t-T) \; .
\end{equation}
The process generalizes in a similar manner for the case of more than one time qubit.

%-----------------------------------------------------------------------------------

\subsection{Two-Qubit Gate Operations}

Two-qubit gate operations may be performed in a straightforward manner.  Thus, a frequency, spatial, or time qubit may be used to control a target time qubit and vice versa.  The details will be omitted here but generalize from the process above and the methods used for spatial qubits.

%%%%%%%%%%%%%%%%%%%%%%%%%%%%%%%%%%%%%%%

\section{Practical Considerations}
\label{sec:practical}

The details of a practical implementation of frequency-based qubit encodings and gate operations have been described elsewhere \cite{LaCour&Ott2015}.  Here we focus on issues specific to spatial and time-based encodings.

We first note that physical implementation of spatial qubits as parallel signals is relatively straightforward, with $m$ spatial qubits requiring $M = 2^m$ pairs of wires, with each pair representing the real and complex parts of $\psi_y$ for each of $M$ values of $y$.  Each $\psi_y$ is, itself, an amplitude modulated signals of $N = 2^n$ frequencies each.

Although we have described these parallel signals as existing on separate wires, this is not the only representation.  The wireless transmission of such signals (described as ``flying qubits'' in the context of quantum communication) may be achieved using distinct spatial modes, for highly collimated signals, or orbital angular momentum multiplexing, for radio-frequency transmissions \cite{Oldoni2015}.  For the purposes of computing, however, we shall consider them to be implemented in hardware as distinct voltage signals.

The most challenging aspect of the implementation is in designing switches to separate out spatial qubit subspaces.  A single-qubit projection of $m$ spatial qubits requires that one identify the half of the $M$ signals corresponding to the 0 subspace for this qubit and the other $M/2$ signals corresponding to the 1 subspace.  This much is straighforward and easily done in a digital controller.  The subspaces will, in general, be interleaved, complicating implementation of the subsequent gate operations.

To address this issue, we propose a three-stage approach.  First, the $M$ signals are reordered through a series of staged swaps (up to $M/2-1$ in all) so that they appear in a standardized ordering (for example, with the address qubit as the most significant bit).  The swaps would be controlled by a digital processor based on the qubit to be addressed.  With this standardized format, each of the $M$ signals may be split and operated upon in a manner similar to that illustrated in Fig.\ \ref{fig:tree2}.  Complex multiplication of the gate would be performed by digitally controlled DC input lines to realize each complex scalar multiplication.  A third and final stage would undo the ordering of the first stage and return the final, transformed state.  Using staged projections, a similar approach may be used to implement two-qubit gates, either between two spatial qubits or between spatial and frequency qubits.  Projections on time-encoded qubits may be performed in a similar manner, with spatial reordering replaced by temporal ordering.

An additional technical issue to address concerns the dynamic range of the signal amplitude.  A typical quantum computing algorithm may transform a state from one in which all amplitude is on a single basis state to one in which the amplitude is distributed evenly across all basis states.  For $m$ spatial and $n$ frequency qubits, this would entail a relative change of scale by a factor of $2^{(m+n)/2}$.  With nominal peak voltages on the order of 1 Volt, this would imply a spread signal well below the noise floor.  To address this, a hybrid analog-digital scheme may be used in which over-unity amplitudes are represented in a digital register yet the mantissa remains as an analog signal \cite{Bryant&al.2012}.

%%%%%%%%%%%%%%%%%%%%%%%%%%%%%%%%%%%%%%%

\color{black}
\section{Applications}
\label{sec:applications}

Previously, we described an application of our quantum emulation approach to the problem of unstructured searches \cite{LaCour&Ostrove2017}.  In particular, for a general $n$-bit Boolean function $f: \{0,1\}^n \mapsto \{0,1\}$, where $f$ may be evaluated in polynomial time, determining the existence of a solution is in the complexity class \textsf{NP} and determining the number of solutions is in the complexity class \#\textsf{P}.

The oracle function $f$ may be represented by an $(n+1)$-qubit unitary operator $U_f$ that may itself be represented by a number of one- and two-qubit gates that scales polynomially with $n$.  If $\ket{x,y}$ is a particular $(n+1)$-qubit basis state, where $x$ is an $n$-bit (input) integer and $y$ is a single (output) bit, then the action of $U_f$ on $\ket{x,y}$ is defined by
\begin{equation}
U_f \ket{x,y} = \ket{x, y \oplus f(x)} \; .
\end{equation}

If we prepare an initial state that is a uniform superposition of $\ket{0,0}, \ldots, \ket{2^n-1,0}$, which in a frequency-encoding scheme reduces to
\begin{equation}
\psi(t) = 2^{n/2} \cos(\omega_nt) \cdots \cos(\omega_1t) \, e^{j\omega_0t} \; ,
\end{equation}
then we may apply $U_f$ to obtain
\begin{equation}
\psi'(t) = 2^{-n/2} \sum_{x=0}^{2^n-1} e^{j\Omega_x t} \exp\left[ (-1)^{f(x)}\omega_ot \right] \; ,
\end{equation}
where $\Omega_x = (-1)^{x_n}\omega_n + \cdots + (-1)^{x_1}\omega_1$.  Using the previously described projection method, this signal may be filtered to produce two projections on qubit 0, namely, the projection corresponding to $f(x) = 0$ and that corresponding to $f(x)=1$.  The latter represents our solution space.  From this projection it is a simple matter to extract the solutions.

We have shown that, even in the presence of noise, for the same number of oracle calls this approach provides a higher probability of success for finding a solution than does an single application of Grover's algorithm on the same device \cite{LaCour&Ostrove2017}.  Furthermore, the time to determine the number of solutions to the $n$-bit Boolean problem scales only linearly with $n$ in our approach, while the time scales as $2^n$ classically and as $\sqrt{2^n}$ quantum mechanically.  Operating in the frequency range of 1 MHz to 1 GHz (corresponding to a mere 10 qubits using frequency encoding), this scaling advantage can provide a speed up of two orders of magnitude against a modern digital processor operating serially \cite{LaCour&Ott&Lanham2017}.

Of course, this speed up is achieved using the exponential resources of increased bandwidth and filter complexity.  Using spatial encoding, however, can provide further capacity without a reduction in speed.  Thus, for example, 1024 spatial modes operating in the aforementioned frequency range would allow for a total 20 fully entangled qubits and a speed up of five orders of magnitude versus a serial processor.  This approach therefore represents not only an improvement over classical serial processing but a better way to perform classical parallel processing, as it relies on representing and manipulating information within the structure of the signal rather than on a mere brute force scaling of computational resources.

\color{black}

%%%%%%%%%%%%%%%%%%%%%%%%%%%%%%%%%%%%%%%

\section{Conclusion}
\label{sec:conclusion}

We have described a scheme whereby a multiqubit quantum state may be represented classically through a collection of parallel, multi-frequency signals.  Within this representation, single- and multi-qubit gate operations may be performed that would allow the realization of any given unitary transformation on this notional state.  In particular, physically distinct signals may become entangled through controlled two-qubit gate operations between the two modal representations.  A similar representation and process may be used for qubits represented in time ordering.

As with any \color{black}known classical emulation of a quantum computer\color{black}, resources scale exponentially with the number of qubits.  Those resources may be bandwidth, physical volume, or temporal length, depending upon the representation.  Using the first two, however, allows for an inherent parallelism that can provide a speedup by several orders of magnitude even under modest scaling.  While quantum states of such scale are easily represented on a single digital processor, transformations of such a state would be significantly accelerated using the approach we have described, and this could provide a distinct, practical computational advantage over current, digital processors.

%%%%%%%%%%%%%%%%%%%%%%%%%%%%%%%%%%%%%%%

\section{Acknowledgment}

The authors would like to thank the Office of Naval Research for their support of this work under Grant No.\ N00014-17-1-2107.

%%%%%%%%%%%%%%%%%%%%%%%%%%%%%%%%%%%%%%%

\bibliographystyle{IEEEtran}
\bibliography{LaCour2018}

% limit of 10 pages, not including the bibliography
\end{document}